\documentstyle[12pt]{article}

\begin{document}

\begin{center}
 {\Large \bf Non-leptonic decays of the $B_c$ into tensor mesons}\\
\vspace{1.3cm}

{\large G. L\'opez Castro$^1$, H. B. Mayorga$^2$ and J. H.
Mu\~noz$^2$}

$^1$ {\it Departamento de F\'{\i}sica, Centro de Investigaci\'on y de
Estudios}\\ {\it Avanzados del IPN, Apdo. Postal 14-740, 07000
M\'exico, D. F., M\'exico} \\

$^2$ {\it Departamento de F\'{\i}sica, Universidad del Tolima,}
{\it A. A. 546, Ibagu\'e, Colombia}

\end{center} 

\vspace{1.3cm} 

\begin{abstract}

We have computed the branching ratios of the exclusive pseudoscalar
(vector) + tensor modes that are allowed in the decays of the $B_c$ meson.
The dominant spectator and annihilation contributions in those decays are
evaluated using the factorization hypothesis.  We find that some of these
decay channels, such as $B_c^- \rightarrow (\rho^-, D_s^{*-},
D_s^{-})\chi_{c2}$ and $B_c^- \rightarrow \pi^- \overline{B_{s2}^{*0}}$,
have branching ratios of the order of $10^{-4}$, which seems to be at the
reach of forthcoming experiments at the LHC.  The inclusive
branching fraction of the  two-body $B_c$ decays involving tensor
particles is approximately $1.28 \times 10^{-3}$. At the
dynamical level, it
is interesting to observe that  the exclusive decays $B_c^- \rightarrow
K^-(\pi^-)\overline{D_2^{*0}}$, $\pi^0 D_2^{*-}$, $\eta^{'} D_{s2}^{*-}$ 
 are dominated by the annihilation contributions. 

\end{abstract}

\vspace{1.3cm}

PACS number(s):  13.25.Hw, 14.40.Nd, 14.40.Lb

\newpage

Studies of the $B_c$ meson decays are important for several reasons. The
mass of the $J^P = 0^-$ meson composed of two heavy quarks
($b\overline{c}$) lies between the masses of the corresponding
$b\overline{q}$ ($q=u$, $d$, $s$) and $b\overline{b}$ systems. While the
light quark in $B_{u,d}$ mesons plays a marginal role as an spectator in
the dominant decays, both $b$ and $\overline{b}$ quarks fully participate
in the decays of beauty
quarkonium states. The beauty-charmed meson share two interesting
features of those systems. Firstly, the phase-space in $B_c$ meson decays
is large enough to allow decays where not only the charm antiquark, but
also the $b$ quark can play the role of spectator. Secondly, in some
decays of the $B_c$ mesons we can expect that
both processes, namely those containing a $c$ quark as spectator (a $b
\rightarrow cW^*$ transition) and the
$b\overline{c} \rightarrow W^*$ annihilation diagrams, play similar 
important roles since they can have  the same  Cabibbo-Kobayashi-Maskawa  
mixing weights. Thus, $B_c$ meson decays can allow a unique place for
testing the interplay of both dynamical processes which are not 
easily accessible in the decays of $B_{u,d}$ mesons.
Furthermore, as also is explored in this paper, some $B_c$ decays have
$W$-annihilation contributions which largely dominate some of the decay
amplitudes. \\ 

  The semileptonic and non-leptonic decays of the $B_c$ meson that are
expected to be the dominant ones, have been considered previously by other
authors \cite{Bc}. Here, we focus on the two-body non-leptonic $B_c$
decays that contain a $J^P=2^+$ tensor meson in the final state, i.e. $B_c
\rightarrow P(V)T$, where $P(V)$ denotes a pseudoscalar (vector) and $T$ a
tensor meson. The analogous decays of the $B_{u,d}$ mesons have been
considered in refs. \cite{herman1,herman2} (see also \cite{btot}). Our
study is exhaustive  in the sense that we make predictions for all the
decays that are allowed by the kinematical constraints and the leading
order dynamical factorizable contributions. As in
our previous works, we have used the non-relativistic quark model of Isgur
{\it et al} (ISGW-model) \cite{isgw} as input for the hadronic matrix
elements $\langle T|J_{\mu}|B_c \rangle$ and $\langle
PT|J_{\mu}|0\rangle $ that are required in our factorization
approximation for the spectator and annihilation contributions,
respectively. Note that factorizable amplitudes proportional to the
matrix element $\langle T| J_{\mu}|0 \rangle$ do not contribute
because this matrix element vanishes identically from Lorentz covariance
considerations.
As a final remark, let us mention that the annihilation amplitudes
will be more important in $B_c$ than in $B_{u}$ decays because they depend
upon the same CKM mixing factors (namely $V_{cb}$) as the spectator
$b \rightarrow c$ contributions to $B_c$ decays. \\

The specific interest of $B \rightarrow XT$  decays is two-fold. First,
we would like to test how quark models work in describing matrix elements
that involve orbital excitations of the $q\overline{q^{'}}$ system (the
tensor meson). Second, we are interested in the evaluation of the
fraction of non-leptonic decays provided by modes containing tensor
mesons. This is important in order to test how different exclusive
channels contribute to the inclusive rates which can in principle be
computed using QCD and the quark-hadron duality \cite{bb}. In addition,
the
study of the $B_c$ decays involving spin-2 (tensor) mesons can provide
complementary tests for the quark models predictions of the orbital pieces
of the meson wavefunctions, which are not accessible with purely lowest
lying $P$ and $V$ states.\\

Our results show that some  branching ratios
of $B_c \rightarrow P(V)T$ decays, 
turn out to be of the order of $10^{-4}$. Those fractions seems to be at
the reach of future experiments at the LHC where it is expected to produce
$2.1 \times 10^{8}$ $B_c$ mesons with an integrated luminosity of 100
fb$^{-1}$ and cuts of $P_T(B_c) > 20$ GeV, $|y(B_c)|<2.5$
\cite{production}. With a further increase in the luminosity one
could reach $10^{10}$ $B_c$ mesons per year
\cite{production1}.  In fact, with a fragmentation ratio of $3.8
\times 10^{-4}$ ($5.4 \times 10 ^{-4}$) for the $B_c$ ($B_c^*$) meson
\cite{braaten}, $10^8$-$10^9$ $B_c$$^{'}$s can be available from a rather
conservative estimate \cite{choudhury}. These samples of $B_c$
mesons, would allow a detailed study of a large diversity of properties
and decay modes of these mesons. \\

Let us start our discussion of $B\rightarrow XT$ decays with the effective
weak Hamiltonian that involve a $b$ quark decay (namely a 
$\bar{c}$ as spectator) \cite{buchalla}:

\begin{equation}
{\cal H}_{eff}(\Delta b=1)= \frac{G_F}{\sqrt{2}}\sum_{q^{'}, Q^{'}, 
q}V_{q^{'}b}V^*_{Q^{'}q}\left[ a_1(\overline{q^{'}}b)(\overline{q}Q^{'}) +
a_2  (\overline{q}b)(\overline{q^{'}}Q^{'})    \right] + h.c.
\end{equation}
The corresponding operator for $B_c \rightarrow XT$ transitions that
proceed through the $c$-antiquark decay ($b$ as spectator) in the $B_c$
meson is given by: 

\begin{equation}
{\cal H}_{eff}(\Delta c=1)= \frac{G_F}{\sqrt{2}}\sum_{Q,
q}V_{cQ}V_{uq}^* \left[  a_1(\overline{Q}c)(\overline{u}q) +
a_2(\overline{u}c)(\overline{Q}q) \right] + h.c.
\end{equation}
In the above expressions $(\overline{q}q^{'})$ is used for  
the $V-A$ current, $Q^{'}$, $q^{'}$ $=u,$ $c$ and  $Q$, $q$ $=d$, $s$,
$G_F$ denotes the Fermi constant,  $V_{ij}$ are the CKM
mixing factors, and  $a_{1,2}$ denote the QCD coefficients. The
effective Hamiltonian that give rises to annihilation amplitudes would
be considered below (see Eq. (4)).\\

 The contribution from $a_1$ corresponds to the $W$-external emission
diagram at the tree level ({\it class I} decays \cite{neubert}) and the
contribution from $W$-internal emission ({\it class II} decays
\cite{neubert}) decays is given by $a_2$.  We also distinguish two
types of suppressed decays according to the occurrence of the CKM factors 
\cite{herman2}: in the {\it type-I(II)}
decays the suppression occurs in the vertex where the boson $W$ is
produced (annihilated). It is important to note that for the
$b$-spectator $B_c$ decays these two kinds of suppression are at the
same order in contrast to the $c$-spectator $B_c$ decays where the
type-I suppression is stronger than in the type-II case. \\

In order to provide numerical values of the branching ratios we use 
the expressions (9) and (11) for the decay rates given in Ref.
\cite{herman1} and the following values of
the CKM elements \cite{pdg}: $|V_{cb}|= 0.0402$, $|V_{ud}|=0.9735$, $|
V_{cs}|=0.9749$, $| V_{us}|=0.2196$, $| V_{cd}|=0.224$, and
$|V_{ub}|=3.3 \times 10^{-3}$. The values for the lifetime and the
mass of $B_c$ were taken from the experimental measurements of
Ref. \cite{cdf} and the value of the masses of
the tensor mesons $B_2^*$ and $B_{s2}^*$ are  5.733  and 5.844 GeV,
respectively \cite{ebert}. All the other masses
required are taken from \cite{pdg}. The values used for the QCD
coefficients are $a_1 = 1.132$ \cite{0103036} and $a_2= 0.29$ (which
is obtained using the relation $|a_2/a_1|=0.26$ of Ref.
\cite{browder}).\\

The decay constants of pseudoscalar mesons $f_P$ (in GeV units)  have the
following central values: $f_{\pi^-} = 0.131$ \cite{pdg}, $f_{\pi^0}
= 0.130$ \cite{pdg}, $f_{\eta} = 0.131$ \cite{neubert},
$f_{\eta^{'}}=0.118$ \cite{neubert}, $f_{D_s} = 0.280$ \cite{cleo},
$f_D = 0.252$, $f_{\eta_c} = 0.393$ \cite{sharma1}, and $f_K = 0.159$ 
\cite{pdg}. The decay constant $f_D$ was obtained using the theoretical
prediction
$f_D/f_{D_s}= 0.90$ \cite{bernard} and the value for $f_{D_s}$ quoted
above. On
the other hand, the central values for the dimensionless decay
constants of vector mesons $f_V$ are \cite{sharma1} $f_{\rho}=0.281$,
$f_{\omega}=0.249$, $f_{\phi}=0.232$, $f_{D_s^*}= 0.128$,
$f_{D^*}=0.124$, $f_{J/\psi}=0.1307$, and $f_{K^*}=0.248$. \\

We have also used the following expressions for the physical states
of the mixing between octet and singlet states of SU(3) with $I=0$:

\begin{eqnarray}
\eta &=& \frac{1}{\sqrt{2}}(u \overline{u} + d \overline{d}) 
\sin\phi_P - (s\overline{s})\cos\phi_P, \nonumber \\
\eta^{'} &=& \frac{1}{\sqrt{2}}(u \overline{u} + d \overline{d}) 
\cos\phi_P + (s\overline{s})\sin\phi_P, \nonumber \\
\omega &=& \frac{1}{\sqrt{2}}(u \overline{u} + d \overline{d}) 
\cos\phi_V + (s\overline{s})\sin\phi_V, \nonumber \\
\phi &=&  \frac{1}{\sqrt{2}}(u \overline{u} + d \overline{d}) 
\sin\phi_V - (s\overline{s})\cos\phi_V, 
\end{eqnarray}
where the mixing angle is given by $\phi_i = \arctan{(1/\sqrt{2})} -
\theta_i$ ($i=P$ or $V$) and the experimental values of $\theta_i$ are 
given by $-20^0$ and $39^0$  \cite{pdg} for pseudoscalar ($\eta,
\eta'$) and  vector ($\omega,\phi$)  mesons, respectively.\\

   With the above convention and numerical values\footnote{Since we are
interested in getting estimated values for the branching fractions, we do
not 
quote their corresponding error bars which at present are dominated by
the uncertainties in the mass and lifetime of the $B_c$ meson.},
we have computed the decay amplitudes for $B_c \rightarrow XT$ modes. 
They are shown in the second column of Tables 1 and 2 (second column
in Table 3),
respectively, for the $PT$ and $VT$ channels when the $c$-antiquark (the
$b$-quark) plays the role of spectator in the $B_c$ decay.
 We do not calculate the $b-$spectator $B_c \rightarrow VT$ decays
because they are completely forbidden by kinematics. We can observe
that the amplitudes of $B_c \rightarrow P(V)T$ are proportional to
only one of the QCD coefficients $a_i$ at the time . The expressions for
the factors  
${\cal F}^{i \rightarrow f}$ and
${\cal F}^{i \rightarrow f}_{\mu\nu}$ in the amplitudes and the properties
of the symmetric polarization tensor $\epsilon_{\mu\nu}$ describing
the tensor meson can be found in Ref. \cite{herman1}.  \\

We  have computed the branching ratios of $B_c \rightarrow PT$ and $B_c
\rightarrow VT$ (last column in tables 1--3) 
using the non-relativistic constituent quark model ISGW \cite{isgw},
which can be reliable to describe the $B_c$ meson composed of
two different heavy quarks \cite{prelovsek}. We restrict ourselves to
this model because the predictions for the required form factors are
not available in a systematic way in other quark model approaches. The
branching ratios obtained for the $B_c^- \rightarrow (\rho^-, D_s^{*-},
D_s^-)\chi_{c2}$ and $B_c^+ \rightarrow \pi^+B_{s2}^{*0}$ decays (which
are proportional to the QCD coefficient
$a_1$) turn out to be of the order of $10^{-4}$, which seems to be at
the reach of future $B_c$ samples at the LHC \cite{production}. \\

The decays $B_c \rightarrow P(V)T$ with the largest branching fractions 
are of the same order that some $B_c \rightarrow PP,
PV, VV$ modes \cite{Bc}.
 Note that the decay $B_c^+ \rightarrow \pi^+
B_{s2}^{*0}$, which proceeds having the $b$-quark as an spectator, 
has a similar branching ratio than $B_c^- \rightarrow (\rho^-, D_s^{*-},
D_s^{-})\chi_{c2}$ 
 because it is favored by the  CKM factor although it is suppressed by
phase
space. As a matter of fact,  if this decay and someone of $B_c^-
\rightarrow (\rho^-, D_s^{*-}, D_s^{-})\chi_{c2}$  were measured,
they would provide a good test for the flavor independence of the QCD
coefficient $a_1$ since in $B_c$ decays with $b$-
or $c$ quarks, they are assumed to be the same. \\

Let us mention that some of  the branching ratios of the {\it type-I}
suppressed decays $B_c \rightarrow VT$ (rows 5-10 in table 2) turn
out to be of the same order than $B(B_c^+ \rightarrow
\rho^+(K^{*+})\gamma)$ ($\approx 10^{-8}-10^{-9}$) \cite{aliev}. 
Observe also  that the ratio $B(B_c \rightarrow VT)/B(B_c 
\rightarrow PT) \approx 3$ when the contribution arises from the $a_1$
coefficient, and
$B(B_c^- \rightarrow D^{*0}D_2^{*-})/B(B_c^- \rightarrow
\overline{D^{*0}}D_2^{*-})$ is different from unity (see the second ant
the last
columns in table 2) because they arise from the transitions $b
\rightarrow c$ and $b \rightarrow u$, respectively. \\

In a recent paper \cite{0103036}, the authors have computed all the
$c$-spectator modes $B_c \rightarrow P(V)\chi_{c2}$ using 
the so-called generalized instantaneous approximation \cite{chang}. In
order to compare their results and ours we show their numerical values
within parenthesis in the third column of tables 1 and 2. Our
results in these decays are in some cases of the same order of  magnitude
than the results of Ref. \cite{0103036} and in the other cases are
smaller by an order of magnitude except in the decay $B_c^- \rightarrow
K^{*-}\chi_{c2}$. \\

Now   we will focus on the study of  the annihilation contributions
to $B_c \rightarrow PT$ decays. The evaluation of the corresponding
contributions to $B_c \rightarrow VT$ would require the knowledge of the 
$\langle VT|J_{\mu}|0 \rangle$ matrix element, which has not been
computed in the literature. As is well known, the annihilation
amplitudes for the two-body non-leptonic $B_{u,d}$ decays are  
helicity-suppressed \cite{neubert,ali}. In fact, as it has been proved in Refs. 
\cite{xing}, the  $B_{u,d} \rightarrow P_1P_2$ decays that proceed 
only via a single W-exchange or W-annihilation quark diagram have
negligible branching ratios, although some of them can
be at the reach of B-factories. On the other hand, in some exclusive
processes where the non-annihilation contribution are highly suppressed,
the annihilation contributions may become important \cite{ali,bauer}. \\

Ali {\it et al.} \cite{ali} have shown explicitly that the annihilation
amplitude in $B_{u,d} \rightarrow P_1P_2$ is suppressed by a hefty 
factor with respect to the non-annihilation contribution, and have
suggested that in the decays $B_{u,d} \rightarrow PV$, $VV$ the situation
may be different because the annihilation quark-diagram  may
enhance the decay rates. We expect that the annihilation contributions to some exclusive $B_c \rightarrow PT$ decays, where the non-annihilation
contribution is too suppressed, may become more important than in the
$B_{u,d}$ decays. This happens because the $B_c$ meson is composed of two
heavy quarks and the annihilation amplitude is proportional to the CKM
factor $V_{cb}$ instead of $V_{ub}$ as in $B_{u,d}$ decays. \\

The effective weak Hamiltonian for the annihilation contributions to
$B_c \rightarrow PT$ is \cite{buchalla}: 

\begin{equation}
{\cal H}^{eff}_a =
\frac{G_F}{\sqrt{2}}\sum_{Q^{'}q}V_{cb}V_{Q^{'}q}^*a_1
(\overline{c}b)(\overline{q}Q^{'}) + h.c.,
\end{equation}
where the subscript $a$ denotes {\it annihilation}, $Q^{'}=u$, $c$ and
$q=d,$ $s$. 
The $a_1$ contribution is associated to the $W$-annihilation for
charged mesons. It is clear that in this case there is not an $a_2$
contribution corresponding to the $W$-exchange for neutral mesons. \\

The annihilation amplitude can be written as \cite{ali}

\begin{equation}
{\cal M}_a (B_c \rightarrow PT) = i
\frac{G_F}{\sqrt{2}}V_{cb}V_{Q^{'}q}^*a_1 <TP|J_{\mu}|0>_a<0|J^{\mu}|B_c>, 
\end{equation}
where $<0|J^{\mu}|B_c> = if_{B_c}(P_{B_c})^{\mu}$ and $<TP|J_{\mu}|0>_a$
is taken from the ISGW-model \cite{isgw}, using the crossing symmetry of
the $<T|J_{\mu}|P>$ amplitude.  Thus, we write the annihilation amplitude
as:

\begin{equation}
{\cal M}_a (B_c \rightarrow PT) =
\frac{G_F}{\sqrt{2}}V_{cb}V_{Q^{'}q}^*a_1f_{B_c}\epsilon^*_{\mu
\nu}(P_P)^{\mu}(P_P)^{\nu} {\cal F}^{PT}(m^2_{B_c}), 
\end{equation}
where $\epsilon^*_{\mu \nu}$ is the polarization of the tensor meson,
${\cal F}^{PT}(m^2_{B_c}) = k_a + (m_P^2 - m_T^2) (b_+)_a + m^2_{B_c}
(b_-)_a$, with $k_a$, $(b_{\pm})_a$ being the $P\rightarrow T$ form
factors evaluated in $q^2 = m^2_{B_c}$. \\

The   decay rate for those processes  which are produced only by the
annihilation contribution is 

\begin{equation}\label{anni}
\Gamma_a(B_c \rightarrow PT) = |{\cal A}_a|^2 \left(
\frac{m_{B_c}}{m_T} \right)^2\frac{|\vec{P_P}|^5}{12 \pi m_T^2},
\end{equation}
with ${\cal A}_a$ given by

\begin{equation}\nonumber
{\cal A}_a = \frac{G_F}{\sqrt{2}}V_{cb}V_{Q^{'}q}^*a_1f_{B_c}{\cal
F}^{PT}(m_{B_c}^2). \\
\end{equation}
\\

To get an idea about the order of magnitude of the annihilation
contribution to $B_c \rightarrow PT$ we can compare the 
expressions given by the Eq. (9) in Ref. \cite{herman1}, and our Eq.
(\ref{anni}) shown above. If a given decay has $W$-emission ($t$) and
annihilation ($a$) contributions, we get the following ratio among them:

\begin{equation}
{\cal R} \equiv \frac{\Gamma_{t}(B_c \rightarrow PT)}{\Gamma_a(B_c
\rightarrow PT)}=  \left[  \frac{V_1^tV_2^ta_if_P{\cal F}^{B_c \rightarrow
T}(m^2_P)}{V_1^aV_2^aa_1f_{B_c}{\cal F}^{PT}(m^2_{B_c})}  \right]^2.\\
\end{equation}
\\

In table 4 we list the ratio ${\cal R}$ for some exclusive  $B_c
\rightarrow PT$ decays that receive both of these contributions.
In our numerical evaluations we have used the value
$f_{B_c}=0.48$ GeV from Ref. \cite{xing}. If we had assumed that the
form factors for the annihilation and the direct diagrams are of the
same order, we would see that the annihilation contribution becomes much
bigger than the $W$-emission contribution. For example, in the
$B_c^- \rightarrow K^-\overline{D_2^{*0}}$ decay, the annihilation
contribution is almost $10^5$ times the tree level one, and
for the decay $B_c \rightarrow \eta^{'} D_{s2}^{*-}$ it becomes 
approximately $10^6$ times the direct contribution. These hefty factors
arise basically from the CKM factors. Thus, we expect that the
annihilation
contribution in these decays, which have a highly suppressed
non-annihilation amplitude, may become important.\\

In summary, in this paper we have computed the branching ratios of the
two-body non-leptonic $B_c$ decays involving tensor mesons. The
measurement of these decays would provide additional tests for the
quark models used to compute the hadronic matrix elements that involve 
orbital excitations of the $q\bar{q'}$ system (as is the case of tensor
mesons). The
decays with the largest branching ratios --of order $10^{-4}$-- are $B_c^-
\rightarrow (\rho^-, D_s^{*-}, D_s^-)\chi_{C2}$ and $B_c^+ \rightarrow
\pi^+ B_{S2}^{*0}$, which proceed dominantly through $c$ and
$b$ quarks as spectators, respectively. The fractions for these decays 
seems to be at the reach of future experiments at the LHC.
On the other hand, we have found that  annihilation contributions to $B_c
\rightarrow PT$ decays may become important in the
exclusive channels $B_c^- \rightarrow K^-(\pi^-)\overline{D_2^{*0}}$,
$\pi^0 D_2^{*-}$, $\eta^{'} D_{s2}^{*-}$,  which have highly 
suppressed non-annihilation contributions. Taking into account all the
results for the estimated exclusive modes given in Tables 1-3, we observe
that the
inclusive production of tensor mesons in $B_c$ meson decays have a
branching ratio of $B(B_c \rightarrow XT)=1.28 \times 10^{-3}$. This
result is dominated by the modes having a $c$-quark as an spectator.\\

{\it Acknowledgements} H. B. Mayorga and J. H. Mu\~noz are grateful
to {\it Comit\'e Central de Investigaciones} (CCI) of the University
of Tolima and COLCIENCIAS for financial support. JHM also thanks to the Physics
Department at Cinvestav for the hospitality while this work was
ended. GLC acknowledges financial support from Conacyt and SNI (M\'exico).

\newpage

\begin{center}
\begin{tabular}{||c|c|c||}
\hline
 Process & Amplitude  & $B(B_c \rightarrow PT)$
\\
\hline\hline
$B_c^- \rightarrow \pi^- \chi_{c2}$ & $V_{cb}V_{ud}^*a_1  f_{\pi^-} {\cal F}^{B_c \rightarrow \chi_{c2}}(m^2_{\pi^-})$ & $7.5 \times 10^{-5}$ $(2.48 \times 10^{-4})$ \\
\hline 
$B_c^- \rightarrow D^0 D_2^{*-}$ & $V_{cb}V_{ud}^*a_2 f_{D^0}{\cal F}^{B_c \rightarrow D_2^*}(m^2_{D^0})$ & $6.26 \times 10^{-8}$\\
\hline 
$B_c^- \rightarrow D_s^- \chi_{c2}$ & $V_{cb}V_{cs}^*a_1 f_{D_s^-}{\cal F}^{B_c \rightarrow \chi_{c2}}(m^2_{D_s^-})$ & $1.54 \times 10^{-4}$ $(4.54\times 10^{-4})$\\ 
\hline
$B_c^- \rightarrow \eta_c D_{s2}^{*-}$ & $V_{cb}V_{cs}^*a_2 f_{\eta_c}{\cal F}^{B_c \rightarrow D_{s2}^*}(m^2_{\eta_c})$ & $1.4 \times 10^{-6}$\\

\hline
$B_c^- \rightarrow \pi^- \overline{D_2^{*0}}$ & $V_{ub}V_{ud}^*a_1 f_{\pi^-}{\cal F}^{B_c \rightarrow D_2^{*0}}(m^2_{\pi^-})$ & $1.79 \times 10^{-9}$\\
\hline
$B_c^- \rightarrow \pi^0 D_2^{*-}$ & $V_{ub}V_{ud}^*a_2 f_{\pi^0}{\cal F}^{B_c \rightarrow D_2^{*-}}(m^2_{\pi^0})/\sqrt{2}$ & $5.81 \times 10^{-11}$ \\
\hline
$B_c^- \rightarrow \eta D_2^{*-}$ & $V_{ub}V_{ud}^*a_2 f_{\eta}\sin{\phi_P}{\cal F}^{B_c \rightarrow D_2^{*-}}(m^2_{\eta})/\sqrt{2}$ & $7.46 \times 10^{-12}$ \\
\hline
$B_c^- \rightarrow \eta^{'} D_2^{*-}$ & $V_{ub}V_{ud}^*a_2 f_{\eta^{'}}\cos{\phi_P}{\cal F}^{B_c \rightarrow D_2^{*-}}(m^2_{\eta^{'}})/\sqrt{2}$ & $5.4 \times 10^{-11}$ \\
\hline
$B_c^- \rightarrow D_s^- \overline{D_2^{*0}}$ & $V_{ub}V_{cs}^*a_1 f_{D_s^-}{\cal F}^{B_c \rightarrow D_2^{*0}}(m^2_{D_s^-})$ & $2.18 \times 10^{-8}$ \\
\hline
$B_c^- \rightarrow \overline{D^0} D_{s2}^{*-}$ & $V_{ub}V_{cs}^*a_2 f_{D^0}{\cal F}^{B_c \rightarrow D_{s2}^{*-}}(m^2_{D^0})$ & $5.66 \times 10^{-9}$ \\
\hline
$B_c^- \rightarrow K^- \chi_{c2}$ & $V_{cb}V_{us}^*a_1 f_{K^-}{\cal F}^{B_c \rightarrow \chi_{c2}}(m^2_{K^-})$ & $5.49 \times 10^{-6}$ $(1.78\times 10^{-6})$ \\
\hline
$B_c^- \rightarrow D^0  D_{s2}^{*-}$ & $V_{cb}V_{us}^*a_2 f_{D^0}{\cal F}^{B_c \rightarrow D_{s2}^{*-}}(m^2_{D^0})$ & $1.94 \times 10^{-8}$ \\
\hline
$B_c^- \rightarrow D^- \chi_{c2}$ & $V_{cb}V_{cd}^*a_1 f_{D^-}{\cal F}^{B_c \rightarrow \chi_{c2}}(m^2_{D^-})$ & $7.56 \times 10^{-6}$ $(1.86 \times 10^{-5})$\\

\hline
$B_c^- \rightarrow \eta_c  D_{2}^{*-}$ & $V_{cb}V_{cd}^*a_2 f_{\eta_c}{\cal F}^{B_c \rightarrow D_{2}^{*-}}(m^2_{\eta_c})$ & $1.92 \times 10^{-8}$ \\
\hline
$B_c^- \rightarrow K^-  \overline{D_{2}^{*0}}$ & $V_{ub}V_{us}^*a_1 f_{K^-}{\cal F}^{B_c \rightarrow D_{2}^{*0}}(m^2_{K^-})$ & $1.43 \times 10^{-10}$ \\
\hline
$B_c^- \rightarrow \pi^0  D_{s2}^{*-}$ & $V_{ub}V_{us}^*a_2 f_{\pi^0}{\cal F}^{B_c \rightarrow D_{s2}^{*-}}(m^2_{\pi^0})/\sqrt{2}$ & $1.99 \times 10^{-11}$ \\

\hline
$B_c^- \rightarrow \eta  D_{s2}^{*-}$ & $V_{ub}V_{us}^*a_2 f_{\eta}\sin{\phi_P}{\cal F}^{B_c \rightarrow D_{s2}^{*-}}(m^2_{\eta})/\sqrt{2}$ & $2.51 \times 10^{-12}$ \\
\hline
$B_c^- \rightarrow \eta^{'}  D_{s2}^{*-}$ & $V_{ub}V_{us}^*a_2 f_{\eta^{'}}\cos{\phi_P}{\cal F}^{B_c \rightarrow D_{s2}^{*-}}(m^2_{\eta^{'}})/\sqrt{2}$ & $1.74 \times 10^{-11}$ \\

\hline
$B_c^- \rightarrow D^-  \overline{D_{2}^{*0}}$ & $V_{ub}V_{cd}^*a_1 f_{D^-}{\cal F}^{B_c \rightarrow D_{2}^{*0}}(m^2_{D^-})$ & $8.57 \times 10^{-10}$ \\
\hline
$B_c^- \rightarrow \overline{D^0}  D_{2}^{*-}$ & $V_{ub}V_{cd}^*a_2 f_{D^0}{\cal F}^{B_c \rightarrow D_{2}^{*-}}(m^2_{D^0})$ & $5.6 \times 10^{-11}$ \\
\hline\hline
\end{tabular}

\end{center}
\begin{center}
Table 1. Decay amplitudes and branching ratios for the c-spectator    $B_c \rightarrow PT$ decays (the amplitudes must be multiplied by $(iG_F/\sqrt{2})\varepsilon^*_{\mu
\nu}p_{B_c}^{\mu}p_{B_c}^{\nu}$). The values within parenthesis in the third column are taken from the Ref. \cite{0103036}.

\end{center}

\begin{center}
\begin{tabular}{||c|c|c||}
\hline
 Process & Amplitude  & $B(B_c \rightarrow VT)$
\\

\hline\hline
$B_c^- \rightarrow \rho^- \chi_{c2}$ & $V_{cb}V_{ud}^*a_1  f_{\rho^-}m^2_{\rho^-} {\cal F}^{B_c \rightarrow \chi_{c2}}_{\mu \nu}(m^2_{\rho^-})$ & $2.38 \times 10^{-4}$$(5.18 \times 10^{-4})$ \\

\hline 
$B_c^- \rightarrow D^{*0} D_2^{*-}$ & $V_{cb}V_{ud}^*a_2 f_{D^{*0}}m^2_{D^{*0}}{\cal F}^{B_c \rightarrow D_2^*}_{\mu \nu}(m^2_{D^{*0}})$ & $3.42 \times 10^{-7}$\\
\hline 
$B_c^- \rightarrow D_s^{*-} \chi_{c2}$ & $V_{cb}V_{cs}^*a_1 f_{D_s^{*-}}m^2_{D_s^{*-}}{\cal F}^{B_c \rightarrow \chi_{c2}}_{\mu \nu}(m^2_{D_s^{*-}})$ & $5.25 \times 10^{-4}$ $(2.4 \times 10^{-3})$\\ 
\hline
$B_c^- \rightarrow J/\psi D_{s2}^{*-}$ & $V_{cb}V_{cs}^*a_2 f_{J/\psi}m^2_{J/\psi}{\cal F}^{B_c \rightarrow D_{s2}^*}_{\mu \nu}(m^2_{J/\psi})$ & $2.06 \times 10^{-5}$\\
\hline
$B_c^- \rightarrow \rho^- \overline{D_2^{*0}}$ & $V_{ub}V_{ud}^*a_1 f_{\rho^-}m^2_{\rho^-}{\cal F}^{B_c \rightarrow D_2^{*0}}_{\mu \nu}(m^2_{\rho^-})$ & $7.43 \times 10^{-9}$\\
\hline
$B_c^- \rightarrow \rho^0 D_2^{*-}$ & $V_{ub}V_{ud}^*a_2 f_{\rho^0}m^2_{\rho^0}{\cal F}^{B_c \rightarrow D_2^{*-}}_{\mu \nu}(m^2_{\rho^0})/\sqrt{2}$ & $2.44 \times 10^{-10}$ \\
\hline
$B_c^- \rightarrow \omega D_2^{*-}$ & $V_{ub}V_{ud}^*a_2 f_{\omega}m^2_{\omega}\cos{\phi_V}{\cal F}^{B_c \rightarrow D_2^{*-}}_{\mu \nu}(m^2_{\omega})/\sqrt{2}$ & $1.21 \times 10^{-10}$ \\
\hline
$B_c^- \rightarrow \phi D_2^{*-}$ & $V_{ub}V_{ud}^*a_2 f_{\phi}m^2_{\phi}\sin{\phi_V}{\cal F}^{B_c \rightarrow D_2^{*-}}_{\mu \nu}(m^2_{\phi})/\sqrt{2}$ & $1.55 \times 10^{-10}$ \\

\hline
$B_c^- \rightarrow D_s^{*-} \overline{D_2^{*0}}$ & $V_{ub}V_{cs}^*a_1 f_{D_s^{*-}}m^2_{D_s^{*-}}{\cal F}^{B_c \rightarrow D_2^{*0}}_{\mu \nu}(m^2_{D_s^{*-}})$ & $8.72 \times 10^{-8}$ \\
\hline
$B_c^- \rightarrow \overline{D^{*0}} D_{s2}^{*-}$ & $V_{ub}V_{cs}^*a_2 f_{D^{*0}}m^2_{D^{*0}}{\cal F}^{B_c \rightarrow D_{s2}^{*-}}_{\mu \nu}(m^2_{D^{*0}})$ & $1.89 \times 10^{-8}$ \\
\hline
$B_c^- \rightarrow K^{*-} \chi_{c2}$ & $V_{cb}V_{us}^*a_1 f_{K^{*-}}m^2_{K^{*-}}{\cal F}^{B_c \rightarrow \chi_{c2}}_{\mu \nu}(m^2_{K^{*-}})$ & $1.33 \times 10^{-5}$ $(3.12 \times 10^{-6})$ \\
\hline
$B_c^- \rightarrow D^{*0}  D_{s2}^{*-}$ & $V_{cb}V_{us}^*a_2 f_{D^{*0}}m^2_{D^{*0}}{\cal F}^{B_c \rightarrow D_{s2}^{*-}}_{\mu \nu}(m^2_{D^{*0}})$ & $7.63 \times 10^{-8}$ \\
\hline
$B_c^- \rightarrow D^{*-} \chi_{c2}$ & $V_{cb}V_{cd}^*a_1 f_{D^{*-}}m^2_{D^{*-}}{\cal F}^{B_c \rightarrow \chi_{c2}}_{\mu \nu}(m^2_{D^{*-}})$ & $2.42 \times 10^{-5}$ $(8.66 \times 10^{-5})$\\
\hline
$B_c^- \rightarrow J/\psi  D_{2}^{*-}$ & $V_{cb}V_{cd}^*a_2 f_{J/\psi}m^2_{J/\psi}{\cal F}^{B_c \rightarrow D_{2}^{*-}}_{\mu \nu}(m^2_{J/\psi})$ & $4.22 \times 10^{-7}$ \\
\hline
$B_c^- \rightarrow K^{*-}  \overline{D_{2}^{*0}}$ & $V_{ub}V_{us}^*a_1 f_{K^{*-}}m^2_{K^{*-}}{\cal F}^{B_c \rightarrow D_{2}^{*0}}_{\mu \nu}(m^2_{K^{*-}})$ & $4.52 \times  10^{-10}$ \\
\hline
$B_c^- \rightarrow \rho^0  D_{s2}^{*-}$ & $V_{ub}V_{us}^*a_2 f_{\rho^0}m^2_{\rho^0}{\cal F}^{B_c \rightarrow D_{s2}^{*-}}_{\mu \nu}(m^2_{\rho^0})/\sqrt{2}$ & $7.91 \times 10^{-11}$ \\
\hline
$B_c^- \rightarrow \omega  D_{s2}^{*-}$ & $V_{ub}V_{us}^*a_2 f_{\omega}m^2_{\omega}\cos{\phi_V}{\cal F}^{B_c \rightarrow D_{s2}^{*-}}_{\mu \nu}(m^2_{\omega})/\sqrt{2}$ & $3.94 \times   10^{-11}$ \\
\hline
$B_c^- \rightarrow \phi  D_{s2}^{*-}$ & $V_{ub}V_{us}^*a_2 f_{\phi}m^2_{\phi}\sin{\phi_V}{\cal F}^{B_c \rightarrow D_{s2}^{*-}}_{\mu \nu}(m^2_{\phi})/\sqrt{2}$ & $4.81 \times 10^{-11}$ \\
\hline
$B_c^- \rightarrow D^{*-}  \overline{D_{2}^{*0}}$ & $V_{ub}V_{cd}^*a_1 f_{D^{*-}}m^2_{D^{*-}}{\cal F}^{B_c \rightarrow D_{2}^{*0}}_{\mu \nu}(m^2_{D^{*-}})$ & $3.26 \times 10^{-9}$ \\
\hline
$B_c^- \rightarrow \overline{D^{*0}}  D_{2}^{*-}$ & $V_{ub}V_{cd}^*a_2 f_{D^{*0}}m^2_{D^{*0}}{\cal F}^{B_c \rightarrow D_{2}^{*-}}_{\mu \nu}(m^2_{D^{*0}})$ & $2.12 \times 10^{-10}$ \\
\hline\hline
\end{tabular}
\end{center}
\begin{center}
Table 2. Decay amplitudes and branching ratios for the c-spectator    $B_c
\rightarrow VT$ decays (the
amplitudes must be multiplied by $(G_F/\sqrt{2})\varepsilon^*_{\mu
\nu}$).  The values within parenthesis in the third column are taken from the
Ref. \cite{0103036}. 
\end{center}

\begin{center}
\begin{tabular}{||c|c|c||}
\hline\hline
Process &  Amplitude & $B(B_c \rightarrow PT)$\\
\hline\hline
$B_c^+ \rightarrow \pi^+ B_{s2}^{*0}$ & $V_{cs}V_{ud}a_1f_{\pi^+}{\cal F}^{B_c \rightarrow B_{s2}^{*0}}(m^2_{\pi^+})$ & $2.01 \times 10^{-4}$\\
\hline
$B_c^+ \rightarrow \overline{K^0} B_{2}^{*+}$ &  $V_{cs}V_{ud}a_2f_{\overline{K^0}}{\cal F}^{B_c \rightarrow B_{2}^{*+}}(m^2_{\overline{K^0}})$ & $4.22 \times 10^{-6}$\\
\hline
$B_c^+ \rightarrow \pi^+ B_{2}^{*0}$ &  $V_{cd}V_{ud}a_1f_{\pi^+}{\cal F}^{B_c \rightarrow B_{2}^{*0}}(m^2_{\pi^+})$ & $1.18 \times 10^{-5}$\\
\hline
$B_c^+ \rightarrow \pi^0 B_{2}^{*+}$ &  $V_{cd}V_{ud}a_2f_{\pi^0}{\cal F}^{B_c \rightarrow B_{2}^{*+}}(m^2_{\pi^0})/\sqrt{2}$ & $3.86 \times 10^{-7}$\\
\hline
$B_c^+ \rightarrow K^+ B_{s2}^{*0}$ &   $V_{cs}V_{us}a_1f_{K^+}{\cal F}^{B_c \rightarrow B_{s2}^{*0}}(m^2_{K^+})$ & $5.03 \times 10^{-7}$\\
\hline
$B_c^+ \rightarrow \eta B_{2}^{*+}$ &  $-\cos{\phi_P}V_{cs}V_{us}a_2f_{\eta}{\cal F}^{B_c \rightarrow B_{2}^{*+}}(m^2_{\eta})$ & $6.46 \times 10^{-8}$\\
\hline
$B_c^+ \rightarrow K^+ B_{2}^{*0}$ &  $V_{cd}V_{us}a_1f_{K^+}{\cal F}^{B_c \rightarrow B_{2}^{*0}}(m^2_{K^+})$ & $1.90 \times 10^{-7}$\\
\hline
$B_c^+ \rightarrow K^0 B_{2}^{*+}$ &  $V_{cd}V_{us}a_2f_{K^0}{\cal F}^{B_c \rightarrow B_{2}^{*+}}(m^2_{K^0})$ & $1.19 \times 10^{-8}$\\
\hline\hline
\end{tabular}
\end{center}
\begin{center}
Table 3. Decay amplitudes and branching ratios for  the b-spectator  $B_c \rightarrow PT$ decays.  The amplitudes must be multiplied by $(iG_F/\sqrt{2})\varepsilon^*_{\mu
\nu}p_{B_c}^{\mu}p_{B_c}^{\nu}$. 
\end{center}

\vspace{3 cm}

\begin{center}
\begin{tabular}{||c|c||}
\hline\hline
$B_c \rightarrow PT$ &  ${\cal R}= \Gamma_t/\Gamma_a$ \\
\hline\hline
$B_c^- \rightarrow K^- \overline{D_2^{*0}}$ & $3.8 \times 10^{-5}$\\
\hline

$B_c^- \rightarrow \eta^{'}D_{s2}^{*-}$ & $1.01 \times 10^{-6}$ \\
\hline
$B_c^- \rightarrow \pi^- \overline{D_2^{*0}}$ & $9.8 \times 10^{-3}$ \\
\hline

$B_c^- \rightarrow \pi^0 D_2^{*-}$ & $3.24 \times 10^{-4}$ \\
\hline\hline
\end{tabular}
\end{center}
\begin{center}
Table 4.  Relation between the tree level and annihilation contributions
to $B_c^- \rightarrow PT$. The values in the second column must be multiplied by $\left[{\cal F}^{B_c \rightarrow T}(m^2_{P})/{\cal F}^{PT}(m^2_{B_c})\right]^2$. 

\end{center}

\end{document}